\documentclass[aps,amsfonts,pra,twocolumn,showpacs,floatfix]{revtex4}
\usepackage{epsfig,amsmath,amssymb,bm,epsf,graphics}

\def\ket#1{\vert#1\rangle}
\def\ketbra#1{\vert#1\rangle\langle#1\vert}
\def\ipr#1#2{\langle#1\vert#2\rangle}

\def\Longarrow{\protect\@lra}
\def\@lra{\relbar\joinrel\relbar\joinrel\relbar\joinrel%
          \relbar\joinrel\rightarrow}

\def\coe#1{{({\rm co}E_{#1})}}
\begin{document}
\title{Connecting the geometric measure of entanglement and entanglement witnesses}

\author{Tzu-Chieh Wei and Paul M. Goldbart}
\affiliation{Department of Physics, 
University of Illinois at Urbana-Champaign, 
1110 West Green Street, Urbana, Illinois 61801-3080, U.S.A.}

\date{\today}

\begin{abstract}
The geometric measure of entanglement is an approach to quantifying
entanglement that is based on the Hilbert-space distance (or, equivalently,
angle) between pure states and their best unentangled approximants. An
entanglement
witness is an operator that reveals entanglement for a given entangled state.
A connection is identified between entanglement witnesses and the geometric
measure of entanglement.  This offers a new interpretation of the geometric
measure of entanglement of a state, and renders it experimentally verifiable,
doing so
most readily for states that are pure.
\end{abstract}
\pacs{03.67.Mn, 03.65.Ud}

\maketitle

\noindent
{\it Introduction\/}: 
Entanglement is now recognized as a resource central to much of quantum 
information processing~\cite{NielsenChuang00}. Thus, characterizing and 
quantifying entanglement has emerged as a prominent theme in quantum 
information theory.  Achievements in quantifying mixed-state entanglement 
lie primarily in  bipartite settings~\cite{Horodecki01}.  For 
{\it multipartite\/} mixed states the issue of entanglement evidently 
presents even greater challenges.  

Recent research on quantifying multipartite entanglement has explored a 
geometric approach.  First introduced by Shimony~\cite{Shimony95} in the 
setting of bipartite pure states, this geometric approach has been 
generalized to multipartite settings by Barnum and 
Linden~\cite{BarnumLinden01}, 
and further developed in Ref.~\cite{WeiGoldbart02}.  In the present Paper, 
our aim is to identify a connection between two apparently distinct aspects of 
entanglement: entanglement witnesses~\cite{LewensteinKrausCiracHorodecki00} 
and the geometric approach to entanglement. As entanglement witnesses
are observables, and can, in principle, be measured, the geometric
measure of entanglement thus becomes verifiable experimentally. 

\smallskip\noindent
{\it Geometric measure of entanglement {\rm(GME)}\/}:  
We begin by briefly reviewing the formulation of the GME in the pure-state 
setting.  Let us start by analyzing a multipartite system 
comprising $n$ parts, each of which may have a distinct Hilbert space.  
Consider the general $n$-partite pure state, 
expanded in the local bases $\{|e_{p_i}^{(i)}\}$: 
$|\psi\rangle=\sum_{p_1\cdots p_n}\chi_{p_1p_2\cdots p_n}
|e_{p_1}^{(1)}e_{p_2}^{(2)}\cdots e_{p_n}^{(n)}\rangle$.
As shown in Ref.~\cite{WeiGoldbart02}, the closest separable pure state 
\begin{equation}
\ket{\phi}\equiv\otimes_{i=1}^n|\phi^{(i)}\rangle=\otimes_{i=1}^{n}
\sum_{p_i}c_{p_i}^{(i)}\,|e_{p_i}^{(i)}\rangle
\end{equation}
obeys the equations
\begin{subequations}
\label{eqn:Eigen}
\begin{eqnarray}
\!\!\!\!\!\!\!\sum_{p_1\cdots\widehat{p_i}\cdots p_n}
\chi_{p_1p_2\cdots p_n}^*c_{p_1}^{(1)}\cdots\widehat{c_{p_i}^{(i)}}\cdots c_{p_n}^{(n)}=
\Lambda\,{c_{p_i}^{(i)}}^*, \\ 
\!\!\!\!\!\!\!\!\!\!\sum_{p_1\cdots\widehat{p_i}\cdots p_n}\chi_{p_1p_2\cdots p_n} {c_{p_1}^{(1)}}^*\cdots\widehat{{c_{p_i}^{(i)}}^*}\cdots {c_{p_n}^{(n)}}^*=
\Lambda\,c_{p_i}^{(i)}\,,
\end{eqnarray}
\end{subequations} 
where the eigenvalue $\Lambda\in[-1,1]$ is associated with the Lagrange 
multiplier enforcing the constraint 
$\ipr{\phi}{\phi}\!=\!1$, 
and \,\,$\widehat{}$\,\, denotes exclusion.  
As discussed in Ref.~\cite{WeiGoldbart02}, the spectrum of eigenvalues 
$\{\Lambda\}$ can be interpreted as the cosine of the angle between 
$|\psi\rangle$ and stationary states $\{\ket{\phi}\}$.  Furthermore, the 
largest eigenvalue $\Lambda_{\max}$, which we call the {\it entanglement 
eigenvalue\/}, corresponds to the separable state closest to $\ket{\psi}$.  
An equivalent way to view $\Lambda_{\max}$ is via 
\begin{eqnarray}
\Lambda^2_{\max}(\ket{\psi})&=&\max_{{\rm separable} \ \phi}
||\ipr{\phi}{\psi}||^2\nonumber \\&=&
\max_{{\rm separable} \ \phi}{\rm Tr}
\big(\ketbra{\phi}\,\ketbra{\psi}\big).
\end{eqnarray}
The precise measure of the entanglement of $\ket{\psi}$ 
adopted in Ref.~\cite{WeiGoldbart02} is
$E_{\sin^2}\equiv 1-\Lambda_{\max}^2$.

\smallskip\noindent
{\it Entanglement witness {\rm (EW)}\/}:  
The entanglement witness ${\cal W}$ for an entangled state $\rho$ is
defined to be an operator that is Hermitian and obeys the following 
conditions:\\
 (i)~${\rm Tr}({\cal W}\sigma)\ge 0$ for all separable states $\sigma$, and\\
(ii)~${\rm Tr}({\cal W}\rho)< 0$.\\
Here, we wish to establish a relationship between $\Lambda_{\max}$ for 
the entangled pure state $\ket{\psi}$ and the optimal element of the set 
of entanglement witnesses ${\cal W}$ for $\ket{\psi}$ that have the specific 
form 
\begin{equation}
\label{eqn:classW}
{\cal W}=\lambda^{2}\openone-\ketbra{\psi}, 
\end{equation} 
this set being parametrized by the real, non-negative number $\lambda^{2}$. By 
{\it optimal\/} we mean that, for this specific form of witnesses, 
the value of the \lq\lq detector\rq\rq\ 
${\rm Tr}\big({\cal W}\ketbra{\psi}\big)$ is as negative as can be. 

In order to satisfy condition~(i) we must ensure that, for any 
{\it separable\/} state $\sigma$, we have 
${\rm Tr}\big({\cal W}\sigma\big)\ge 0$.  
As the density matrix for any separable state can be decomposed 
into a mixture of {\it separable pure\/} states 
[i.e., $\sigma=\sum_i\ketbra{\phi_i}$ 
where $\{\ket{\phi_i}\}$ are separable pure states], 
condition~(i) will be satisfied as long as 
${\rm Tr}\big({\cal W}\ketbra{\phi}\big)\ge 0$ 
for all separable {\it pure} states $\ket{\phi}$.
This condition is equivalent to 
\begin{equation}
\lambda^{2} -||\ipr{\psi}{\phi}||^2\ge 0\ 
(\mbox{for all separable}\ \ket{\phi}),
\end{equation}
which leads to 
\begin{equation}
\lambda^{2} \ge
\max_{\ket{\phi}}||\ipr{\psi}{\phi}||^2=
\Lambda^2_{\max}(\ket{\psi}).
\end{equation}

Condition~(ii) requires that 
${\rm Tr}\big({\cal W}\ketbra{\psi}\big)<0$, in order for ${\cal W}$ 
to be a valid EW for $\ket{\psi}$; 
this gives $\lambda^{2} - 1<0$.  
Thus, we have established the range of $\lambda$ for which 
$\lambda^{2}\openone-\ketbra{\psi}$ 
is a valid EW for $\ket{\psi}$: 
\begin{equation}
\Lambda^2_{\max}(\ket{\psi}) \le \lambda^{2} < 1.
\end{equation}

With these preliminaries in place, we can now establish the connection 
we have been seeking.  Of the specific family~(\ref{eqn:classW}) of 
entanglement witnesses for $\ket{\psi}$ that we have been considering, 
the one of the form 
${\cal W}_{\rm opt}=\Lambda^2_{\max}(\ket{\psi})\openone -\ketbra{\psi}$ 
is optimal, in the sense that it achieves the most negative value for 
the detector 
${\rm Tr}\big({\cal W}_{\rm opt}\ketbra{\psi}\big)$: 
\begin{equation}
\min_{{\cal W}}{\rm Tr}\big({\cal W}\ketbra{\psi}\big)=
{\rm Tr}\big({\cal W}_{\rm opt}\ketbra{\psi}\big)= 
-E_{\sin^2}(\ket{\psi}),
\end{equation}
where ${\cal W}$ runs over the class~(\ref{eqn:classW}) 
of witnesses. 

\smallskip\noindent
{\it Some illustrative examples\/}:  
For the state $\ket{\rm GHZ}\equiv(\ket{000}+\ket{111})/{\sqrt{2}}$
the optimal witness is 
\begin{equation}
{\cal W}_{\rm GHZ}=\frac{1}{2}\openone-\ketbra{\rm GHZ}
\end{equation}
and 
${\rm Tr}\big({\cal W}_{\rm GHZ}\ketbra{{\rm GHZ}}\big)=
-E_{\sin^2}(\ket{{\rm GHZ}})=-1/2$. 
Similarly, for the states 
$\ket{W}\equiv(\ket{001}+\ket{010}+\ket{100})/{\sqrt{3}}$ and $\ket{\widetilde{W}}\equiv(\ket{110}+\ket{101}+\ket{011})/{\sqrt{3}}$
we have 
\begin{equation}
{\cal W}_{\rm W}=\frac{4}{9}\openone-\ketbra{W} \ \ \mbox{and} \ \
{\cal W}_{\rm \widetilde{W}}=\frac{4}{9}\openone-\ketbra{\widetilde{W}}
\end{equation}
and 
${\rm Tr}\big({\cal W}_{\rm W}\ketbra{{\rm W}}\big)=
-E_{\sin^2}(\ket{{\rm W}})=-5/9$, and similarly for $\ket{\widetilde{W}}$. 
For the four-qubit state 
$\ket{\Psi}\equiv
(\ket{0011}+\ket{0101}+\ket{0110}+
 \ket{1001}+\ket{1010}+\ket{1100})/\sqrt{6}$
the optimal witness is
\begin{equation}
{\cal W}_{\Psi}=\frac{3}{8}\openone-\ketbra{\Psi}
\end{equation}
and 
${\rm Tr}\big({\cal W}_{\Psi}\ketbra{\Psi}\big)=
-E_{\sin^2}(\ket{\Psi})=-5/8$.
In passing, we note that linear combinations of witnesses---preferably
optimal ones---can be used to detect entanglement for mixed states,
as we shall illustrate later.
We also note that the non-optimal witnesses can also be 
of use, e.g., in classifying and detecting distinct types of entangled 
states; see Ref.~\cite{AcinBrussLewensteinSanpera01}.  
Furthermore, as entanglement witnesses are {\it Hermitian\/} operators, they
can, at least in principle, be 
realized experimentally.  

\begin{figure}[t]
\centerline{\psfig{figure=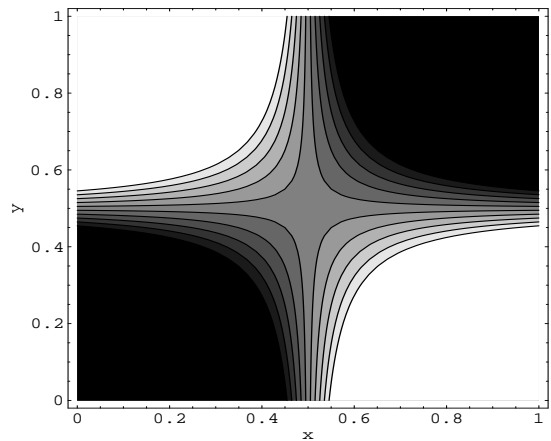,width=7.2cm,height=6cm,angle=0}}
\vspace{- 0.4cm}
\caption{Contour plot of the detector, ${\rm Tr}\big({\cal
W}(y)\rho(x)\big)$. The darker the region, the more negative the detector.
White regions correspond to positive values of the detector, where ${\cal W}$
is not a good entanglement witness, and are therefore left ungraded.}  
\label{fig:witnessWW}
\vspace{-0.2cm}
\end{figure}

\smallskip\noindent
{\it Mixed states\/}: 
We conclude by briefly commenting on the EW/GME connection for mixed states.  
The GME can be generalized to mixed states $\rho$ via the {\it convex 
hull\/} construction (indicated by ``co'').  The essence of this 
construction is a minimization over all decompositions 
$\rho=\sum_i p_i\,|\psi_i\rangle\langle\psi_i|$ 
into pure states:
\begin{eqnarray}
\label{eqn:Emixed}
E(\rho)
\equiv
\coe{\sin^2}(\rho)
\equiv
{\min_{\{p_i,\psi_i\}}}
\sum\nolimits_i p_i \, 
E_{\sin^2}(|\psi_i\rangle).
\end{eqnarray}
By using the analysis for pure states given above, we can rewrite the 
mixed-state entanglement as follows:
\begin{subequations}
\begin{eqnarray}
\label{eqn:Emixed2}
E(\rho)
&\equiv&\phantom{-}
{\min_{\{p_i,\psi_i\}}}
\sum\nolimits_i p_i \, 
E_{\sin^2}(|\psi_i\rangle)
\nonumber \\
&=&-
\max_{\{p_i,\psi_i\}}
\sum\nolimits_i p_i 
\min_{{\cal W}_i}{\rm Tr}\big({\cal W}_{i}\ketbra{\psi_i}\big) \\
&=&-
\max_{\{p_i,\psi_i\}}
\sum\nolimits_i p_i {\rm Tr}\big({\cal W}_{\psi_i}\ketbra{\psi_i}\big),
\end{eqnarray}
\end{subequations}
where ${\cal W}_{\psi_i}$ is the optimal EW corresponding to
$\ket{\psi_i}$.  Said equivalently, one can express the 
entanglement of $\rho$ in terms of the optimal witnesses for the 
pure states that feature in the optimal decomposition.  The 
GME for a mixed state, then, detects the minimum 
average of the \lq\lq fidelities\rq\rq\ between the pure components 
and their optimal witnesses. Although, for any entangled mixed state
$\rho$, there always exists an entanglement witness ${\cal W}$, and
by a trivial rescaling one can then have ${\rm Tr}\big({\cal 
W}\rho\big)=-E(\rho)$,
it would be preferable to have a single, 
simple entanglement witness that is directly connected to the 
entanglement of the mixed state $\rho$. However, we do not know of any simple
one.

Before ending our discussion of mixed states, we 
elaborate on a point made earlier, i.e., that linear combinations
(with non-negative coefficients) of optimal entanglement witnesses
can be used to establish entanglement of mixed states. For illustration, 
consider the following family of mixed states:
\begin{equation}
\label{eqn:rhoX}
\rho(x)\equiv x\ketbra{W}+(1-x)\ketbra{\widetilde{W}},
\end{equation}
the GME of which is calculated analytically in Ref.~\cite{WeiGoldbart02}
and which is entangled for all values of $x\in[0,1]$. We can actually
construct EW's that establish the entanglement of $\rho(x)$.
Consider a linear combination of optimal witnesses of $\ket{\rm W}$
and $\ket{\widetilde{W}}$:
\begin{equation}
{\cal W}(y)\equiv y\,{\cal W}_{\rm W}+(1-y){\cal W}_{\rm \widetilde{W}}\,,
\end{equation}
with $y\in[0,1]$. If, for any given $x$, there exists a value of $y\in[0,1]$
such that ${\rm Tr}\big({\cal W}(y)\rho(x)\big)<0$ then $\rho(x)$ is 
evidently entangled.
Figure~\ref{fig:witnessWW} shows that this is indeed
the case (see captions for details). This illustrates the usefulness of linear
combinations of 
pure-state optimal witnesses. It would be interesting
to know whether witnesses for bound entangled states
be constructed by this approach.

\smallskip\noindent
{\it Concluding remarks\/}: 
Although the observations we have made are, from a technical 
standpoint, elementary, we nevertheless find it intriguing that two distinct
aspects of entanglement---the geometric measure of entanglement and
entanglement witnesses---are so closely related.  Furthermore, this connection sheds 
new light on the content of the geometric measure of entanglement.  In 
particular, as entanglement witnesses are Hermitian operators, they can, at
least in principle, be realized experimentally. Their connection with the
geometric measure of entanglement ensures that the geometric measure of
entanglement can, at least in principle, be verified experimentally.

\noindent
{\it Acknowledgments\/}: 
This work was supported by NSF Award EIA01-21568 and by 
DOE DMS Award DEFG02-91ER45439 through the FS-MRL at UIUC.
TCW acknowledges a Mavis Memorial Fund Scholarship. 


\begin{thebibliography}{99} 
\bibitem{NielsenChuang00}
See, e.g., 
M. Nielsen and I. Chuang,
{\sl Quantum Computation and Quantum Information\/}
(Cambridge University Press, 2000).
\bibitem{Horodecki01}
For a review, see 
M. Horodecki, 
Quant. Info. Comp. {\bf 1\/}, 3 (2001), 
and references therein.
\bibitem{Shimony95}
A. Shimony, 
Ann. N.Y. Acad. Sci. {\bf 755\/}, 675 (1995).
\bibitem{BarnumLinden01}
H. Barnum and N. Linden, 
J. Phys. A: Math. Gen. {\bf 34\/}, 6787 (2001).
\bibitem{WeiGoldbart02}
T. C. Wei and P. M. Goldbart, quant-ph/0212030.
\bibitem{LewensteinKrausCiracHorodecki00} 
M. Lewenstein, B. Kraus, J. I. Cirac, and P. Horodecki,
Phys. Rev. A {\bf 62\/}, 052310 (2000).
\bibitem{AcinBrussLewensteinSanpera01}
A. Ac\'in, D. Bru$\beta$, M. Lewenstein, and A. Sanpera, 
Phys. Rev. Lett. {\bf 87}, 040401 (2001).
\end{thebibliography}
\end{document}